# Valency of rare earths in $R$In$_3$ and $R$Sn$_3$: *Ab initio* analysis of electric-field gradients

S. Jalali Asadabadi,[1] S. Cottenier,[2,*] H. Akbarzadeh,[1] R. Saki,[1] and M. Rots[2]

[1]*Department of Physics, Isfahan University of Technology, Isfahan 84154, Iran*
[2]*Instituut voor Kern- en Stralingsfysica, Katholieke Universiteit Leuven, Celestijnenlaan 200 D, B-3001 Leuven, Belgium*

In $R$In$_3$ and $R$Sn$_3$ the rare earth ($R$) is trivalent, except for Eu and Yb, which are divalent. This was experimentally determined in 1977 by perturbed angular correlation measurements of the electric-field gradient on a $^{111}$Cd impurity. At that time, the data were interpreted using a point charge model, which is now known to be unphysical and unreliable. This makes the valency determination potentially questionable. We revisit these data, and analyze them using *ab initio* calculations of the electric-field gradient. From these calculations, the physical mechanism that is responsible for the influence of the valency on the electric-field gradient is derived. A generally applicable scheme to interpret electric-field gradients is used, which in a transparent way correlates the size of the field gradient with chemical properties of the system.

## I. INTRODUCTION

Many rare earths ($R$) and group IIIa or IVa elements ($X$) form stable $RX_3$ compounds in the AuCu$_3$ structure (Fig. 1). They can be considered as an *f*-element metal homogeneously diluted in an *sp*-element metal, and therefore serve as a good case to study the interaction between *f* and *sp* electrons in the regime of large *f*-electron concentrations. For instance, predictions of the Rudermann-Kittel-Kasuya-Yosida (RKKY) model about the appearance of magnetism on the intrinsically magnetic $R$ atoms immersed in a matrix of nonmagnetic $X$ atoms can be tested.[1] Another question that has attracted much attention is the valency of the rare earth elements in these compounds. In pure rare earth metals they are trivalent, except for Eu and Yb, which are divalent. In compounds too, most rare earths are trivalent, except for Sm, Eu, Tm, and Yb, which appear in a divalent as well as in a trivalent configuration. What will be the rare earth's valency in an $RX_3$ compound? Valencies can be inferred through, e.g., lattice constant measurements: $R^{2+}$ ions are larger than $R^{3+}$, leading to a larger lattice constant in the former case. Whereas this effect is 10% for pure rare earths,[2] it rapidly reduces when $R$ is diluted. For $R=$(Eu, Yb) in $R$Sn$_3$ and $R$In$_3$, a small increase of less than 2% can be seen when compared to other $R$In$_3$ and $R$Sn$_3$ compounds.[3,4] This was taken as an indication for the divalency of Eu and Yb also in these particular series of compounds. The temperature dependence of the susceptibility of Yb in YbIn$_3$ supported this assignment.[5] The question was finally settled when Schwartz and Shirley[4] in 1977 measured the electric-field gradient (EFG; see Sec. II A) at a $^{111}$Cd impurity in $R$In$_3$ and $R$Sn$_3$. For $R=$(Eu, Yb) the electric-field gradient showed a dramatic drop of 50% relative to other rare earths, which by a simple point charge model (see the Appendix) could be related to a changing valency. The effect was so striking that the Schwartz and Shirley (SS) measurements are considered as the archetypical example of valency determination by electric-field gradients ever since.[6,7]

A weak point in the analysis of SS is their reliance on a point charge model. It was the only model for EFG interpretation at that time, but much more sophisticated interpretation schemes based on *ab initio* methods are available nowadays (see, e.g., Refs. 8–13). It is the goal of this paper to revisit the SS data using an *ab initio* electronic structure method. We will examine whether the same conclusion about the valencies can be reached, and we will show how this way of analysis significantly increases insight about the EFG. It will be pointed out how this new understanding can be used to solve questions in related actinide compounds.

## II. FORMULATION OF THE PROBLEM

### A. The EFG at $X$ in $RX_3$

In this paper, we will deal with the main component $V_{zz}$ of the EFG tensor. The electric-field gradient tensor is a symmetric traceless tensor of rank 2 (five independent components), formed by the second derivatives of the electric potential due to the electrons, evaluated at the nucleus. The physical interpretation of its main component $V_{zz}$ is that it is proportional to the deviation from cubic symmetry of the (valence) electron charge distribution in the near vicinity ($\leq 0.2$ Å) of a particular nucleus $X_0$, which is either a regular constituent of the solid or a highly diluted impurity ($V_{zz}=0$ means a charge distribution with cubic or higher symmetry). The value of $V_{zz}$ is determined by the chemistry of the first few atoms surrounding $X_0$. The AuCu$_3$ structure is shown in Fig. 1, in an unconventional setting with an atom X at the center of the cell. The 12 other atoms in Fig. 1 (four $R$ atoms and eight $X$ atoms) are the 12 nearest neighbors of $X$. The point symmetry at $X$ is $4/mmm$, which is lower than cubic: $V_{zz}$ on $X$ will be different from zero. If in the entire crystal a single $X$ atom is replaced by some other atom $X_0$, the point symmetry at this $X_0$ site will remain $4/mmm$. The value of $V_{zz}$ will change however: the interaction $(R,X) \leftrightarrow X_0$ is different from the original interaction $(R,X) \leftrightarrow X$, which changes the size (but not the symmetry) of the charge distribution near $X_0$. In reality, we will replace more than one atom $X$ by an impurity $X_0$, but the concentration of $X_0$ will be low enough to prevent interactions between different impurities. In experimental conditions this means that one can introduce impurities $X_0$ at a ppm concentration into $RX_3$, measure $V_{zz}$ at $X_0$ (which will be different from $V_{zz}$ at $X$), and still learn something that is valid for the pure $RX_3$ compound. Experimental methods that can determine $V_{zz}$ are Mössbauer spectroscopy (MS), perturbed angular correlation spectroscopy (PAC), nuclear quadrupole resonance spectroscopy (NQR), and others.[14] $V_{zz}$ cannot be measured directly. The quantity that is experimentally accessible is the *electric hyperfine splitting* $\Delta E_{\text{hf}}$: a splitting in the electronic and nuclear energy levels that is proportional to the product of $V_{zz}$ and the nuclear quadrupole moment $Q$. If for the nucleus $X_0$ we know $Q$ from nuclear physics, then the condensed matter property $V_{zz}$ can be obtained from $\Delta E_{\text{hf}}$ in the following way:

$$\Delta E_{\text{hf}} = f(I) e Q V_{zz}, \quad (1)$$

with $e$ the electron charge and $f(I)$ a known function of the nuclear spin. $Q$ is measured in barns (1 b = $10^{-28}$ m$^2$) and $V_{zz}$ in V/m$^2$. In this paper all $V_{zz}$—whether they are calculated or measured—are given in units of $10^{21}$ V/m$^2$. They can be converted to $^{119}$Sn Mössbauer splittings $eQV_{zz}/2$ (mm/s) by multiplying with 0.124 914 and in $^{111}$Cd PAC frequencies $eQV_{zz}/h$ (MHz) by multiplying with 20.0693.

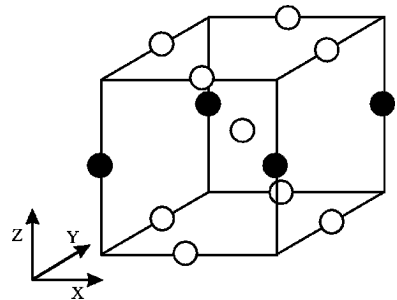

FIG. 1. The $RX_3$ structure in a less traditional setting with $X$ at the center of the unit cell, showing the local symmetry of the $X$ site. Black $= R$, white $= X$.

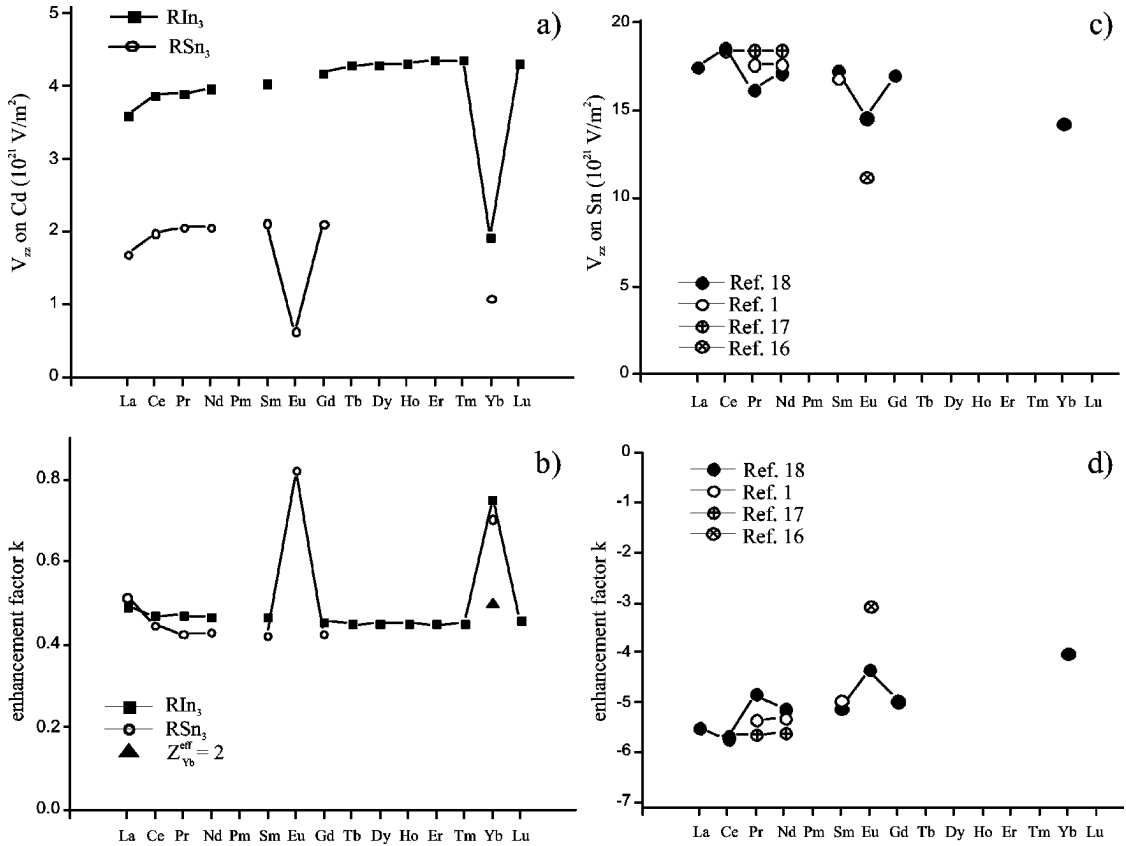

FIG. 2. (a) Experimental $V_{zz}$ on Cd in $R$In$_3$ and $R$Sn$_3$ (Ref. 4). (b) Electronic enhancement factor $k$ derived from (a), if $Z_R^{\text{eff}}=3$ and $Z_{(\text{In,Sn})}^{\text{eff}}=(1,2)$ is used. The value for $Z_R^{\text{eff}}=2$ is indicated too. (c) Experimental $V_{zz}$ on Sn in $R$Sn$_3$ (Refs. 1, 16–18). (d) The electronic enhancement factor $k$ derived from (c), assuming $Z_R^{\text{eff}}=3$ and $Z_{\text{Sn}}^{\text{eff}}=2$.

### B. A point charge model analysis

Using the perturbed angular correlation (PAC) method, SS in 1977 measured $V_{zz}$ at a $^{111}$Cd impurity ($=X_0$) in as many $R$In$_3$ and $R$Sn$_3$ compounds as they were able to produce.[4] Their results are shown in Fig. 2(a). The large drops at Eu and Yb (EuIn$_3$ could not be produced) are immediately visible. These results were analyzed using a point charge model (PCM; see the Appendix for a detailed description). SS used a value $Q=0.44$ b for the $^{111}$Cd quadrupole moment, which is very different from the value $Q=0.83$ b known today.[15] This invalidates their original PCM analysis, which was based on a fortunate but accidental numerical agreement with experiment. Using the correct value of the quadrupole moment, their line of thought can be reformulated as follows.

If a reasonable choice is made for the effective charges, say $Z_R^{\text{eff}}=3$ ("trivalent") and $Z_{\text{In}}^{\text{eff}}=1$, the electronic enhance-ment factor $k$ can be extracted from the measured $V_{zz}$. This is shown in Fig. 2(b), based on the SS data. As the chemical properties of all lanthanides are quite similar, one does not expect $k$ to vary very much as a function of $R$. This is clearly the case in Fig. 2(b), except for $R=$ (Eu, Yb). Such a jump in $k$ is telltale for a sudden change in chemistry, for instance a change in valency. If one looks to the value of $V_{zz}$ in Fig. 2(a), one sees that it drops by a factor of 2 for YbIn$_3$. This is consistent with $Z_{Yb}^{eff}=2$ ("divalent"), such that $Z_{Yb}^{eff}-Z_{In}^{eff}=1$ is only half as large as $Z_R^{eff}-Z_{In}^{eff}=2$. Moreover, if $k$ is determined using $Z_{Yb}^{eff}=2$, then this $k$ more or less fits in the trend of other rare earths [triangle in Fig. 2(b)], which could be taken as additional support for divalent Yb. Similar $k$ anomalies and drops in $V_{zz}$ can be seen for Eu and Yb in the $R$Sn$_3$ series. If an effective charge $Z_{Sn}^{eff}=2$ is used, it can be understood why $V_{zz}$ for the $R$In$_3$ series is on average twice as large as that for the $R$Sn$_3$ series: $Z_R^{eff}-Z_{In}^{eff}=2$ is twice as much as $Z_R^{eff}-Z_{Sn}^{eff}=1$. Determination of $k$ with $Z_{Eu/Yb}^{eff}=2$ is not possible here, as it leads to a division by 0.

### C. Shortcomings of the PCM

The preceding section describes some successes of the PCM to understand the valencies in these compounds. Curiously enough, these successes come together with some serious failures as well. For instance, $V_{zz}$ for Cd in EuSn$_3$ and YbSn$_3$ would be expected to be zero: $Z_{Eu}^{eff}-Z_{Sn}^{eff}=0$. Figure 2(a) shows that this is not the case. Even worse, the PCM completely fails when it is used to explain the EFG on Sn in the $R$Sn$_3$ series. Measuring $V_{zz}$ in such a case can be done with Mössbauer spectroscopy on $^{119}$Sn.[1,16–18] No impurities are needed here, and one would therefore expect an even better-defined and easier situation for the PCM. In Fig. 2(c) $V_{zz}$ on Sn appears to be much larger than $V_{zz}$ on Cd, which gives hope to observe a drop of much larger absolute magnitude. If the trivalent case $Z_R^{eff}-Z_{Sn}^{eff}=1$ corresponds to about $18 \times 10^{21}$ V/m$^2$, one would expect that for EuSn$_3$ with $Z_{Eu}^{eff}-Z_{Sn}^{eff}=0$ $V_{zz}$ is zero or at least rather small. The data in Fig. 2(c) show that with some optimism a drop can be observed for both Eu and Yb (it depends on the particular measurement, the data of Sanchez et al.[18] are best documented and most systematic). But it is certainly not a large drop, and also the enhancement factor in Fig. 2(d) is not too seriously affected.

It is not clear why the PCM appears to give good insight about some aspects of the problem, but badly fails for very related aspects. In such a case, can one trust the "insight" that the PCM offers there where it works? Or is the agreement with experiment just good luck? After all, why should the effective charges $Z_R^{eff}$ be identical to the concept of a valency (see Sec. IV C for an exact definition of the latter)? Schwartz and Shirley wrote: "We will not attempt a quantitative interpretation of the value of $V_{zz}$, which would require a rather elaborate calculation of dubious value in light of the present understanding of the contributions to electric-field gradients in metals and alloys." This "present understanding" has much improved by now, and giving a quantitative interpretation of $V_{zz}$ by modern *ab initio* methods is exactly what we will do in the remainder of this paper.

### D. Questions to answer

In the preceding sections, we sketched several questions that our *ab initio* study should answer. They can be summarized as follows. (1) Why is $V_{zz}$ at Cd in $R$In$_3$ and $R$Sn$_3$ strongly reduced for $R=$ (Eu, Yb)? (2) Why is this not (or at least much less) the case for $V_{zz}$ at Sn? (3) Why is $V_{zz}$ at Cd in these compounds much smaller than $V_{zz}$ at Sn, and more generally, how can we understand the size of $V_{zz}$ at the $4/mmm$ site in these compounds? (4) Why is $V_{zz}$ at Cd in $R$Sn$_3$ half as large as in $R$In$_3$ if $R \neq$ (Eu, Yb)?

In order to tackle these questions, we will use a visualization tool—the *anisotropy function* $\Delta p(E)$—that allows for a transparent interpretation of the qualitative behavior of $V_{zz}$.

### III. COMPUTATIONAL DETAILS

Within density functional theory (DFT), the full-potential linearized augmented plane wave (FLAPW) method as implemented in the WIEN code[19] was used to solve the Kohn-Sham equations. For all calculations reported here, the generalized gradient approximation[20] (GGA) for the exchange-correlation functional was used. The EFG on $X$ in $RX_3$ was calculated using the crystallographic unit cell for $RX_3$ as shown in Fig. 1 (contains four atoms). If an impurity $X_0$ is introduced, the cell is first doubled in all three directions such that the new supercell contains 32 atoms. Then the origin is shifted to an $X$ atom, and the $X$ atom at the new origin and those at the new corners of the supercube are replaced by $X_0$. This supercell has bcc symmetry, and can be represented by a primitive cell with 16 atoms (1 impurity $X_0$, 4 $R$, and 11 $X$ atoms), which is our actual supercell used for the calculations. As we cannot take into account strong correlations very accurately (see Sec. IV C), we will not aim for absolute accuracy of the calculated $V_{zz}$. We therefore adopted an average lattice constant of 4.6418 Å for all compounds (which deviates at most 0.1 Å from the experimental lattice constants) and did not allow for possible relaxation of the positions of atoms surrounding $X_0$. It was checked for one example (Cd in SmIn$_3$) that a change in the lattice constant of 0.1 Å changes $V_{zz}$ on Cd by 10%. Taking into account relaxation of the neighbors of Cd changes the nearest-neighbor distance by 1% and $V_{zz}$ by 3%. This shows that our choice of a constant and unrelaxed lattice yields an accuracy on $V_{zz}$ of 10%, which is sufficient for our purposes. Moreover, this approach allows us to attribute all changes in $V_{zz}$ to the chemistry of the compound and not to the size of the unit cell, which allows for a clearer determination of the physics at work. For the same reason of not aiming for absolute accuracy, we adopted rather low requirements for computational precision. In the FLAPW procedure wave functions, charge density, and potential are expanded in spherical harmonics within nonoverlapping atomic spheres of radius $R_{MT}$ and in plane waves in the remaining space of the unit cell. $R_{MT}$ values of 2.5 bohr were chosen for $X$ in the small unit cell and for Cd in the supercell. $R_{MT}$ for $X$ in the supercell and for R in all cases was taken to be 2.65 bohr. The maximum $l$ for the waves inside the atomic spheres was confined to $l_{max}=10$. The wave functions in the interstitial region were expanded in plane waves with a cutoff of $k_{max}=8/R_{MT}^{min}$ for the small cell and $k_{max}=7/R_{MT}^{min}$ for the supercell. The charge density was Fourier expanded up to $G_{max}=14$. A mesh of 165 special $k$ points was taken in the irreducible wedge of the Brillouin zone for the small cell, and 75 for the supercell. These relatively low requirements enabled us to calculate many different cases with nevertheless limited computer resources. The numerical accuracy on $V_{zz}$ was checked to be better than 5%, and the Density Of States (which will be shown in Sec. IV to be the key ingredient of the explanation) did not change any more when going to stronger requirements.

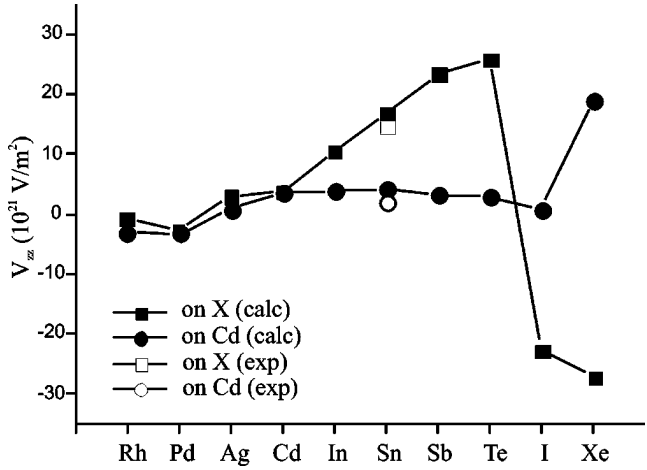

FIG. 3. Solid symbols: calculated $V_{zz}$ on X and calculated $V_{zz}$ on a Cd impurity in a series of mostly hypothetical Eu$X_3$ compounds ($X=$Rh $\rightarrow$Xe). Open symbols: experimental values for $X=$Sn.

## IV. EFG CALCULATIONS IN $RX_3$

### A. $V_{zz}$ at X in Eu$X_3$

As a first step, we want to understand why in the impurity-free series $R$Sn$_3$, $V_{zz}$ at Sn has the particular value of about $18\times10^{21}$ V/m$^2$ [see Fig. 2(c)]. To that end, in Fig. 3 $V_{zz}$ is calculated at X in Eu$X_3$, for $X=$Rh$\rightarrow$Xe ("regular" GGA is used, as opposed to the open core scheme that will be applied in Sec. IV C). We do not care whether our fixed lattice constant of 4.6418 Å is the right one for these compounds, or even whether these compounds exist at all (probably most of them will not). What matters is understanding the physical mechanism that determines the value of $V_{zz}$ in such structures. Figure 3 shows that $V_{zz}$ is close to zero for transition metals (Rh to Cd), steadily increases up to Te and then suddenly changes to large and negative values for I and Xe. One of the values—$V_{zz}$ on Sn in EuSn$_3$—can be compared with experiment [from Fig. 2(c)] and agrees with it. In Sec. II A it was told that $V_{zz}$ expresses the deviation from spherical symmetry of the electron density in the immediate environment of X. In Ref. 10 an explicit expression of this asphericity is given in terms of $p$ and $d$ orbitals ($s$ orbitals have intrinsic spherical symmetry). We will see soon that we need only the $p$ orbitals here, and for them this expression reads

$$V_{zz}=V_{zz}^p+V_{zz}^d, \qquad (2)$$

$$V_{zz}^p=\Delta p(E_F)\left\langle\frac{1}{r^3}\right\rangle_p, \qquad (3)$$

$$\Delta p(E_F)=\frac{1}{2}p_{xy}^I(E_F)-p_z^I(E_F), \qquad (4)$$

$$p_i^I(E_1)=\int_{-\infty}^{E_1}p_i(E)dE. \qquad (5)$$

Here $\langle 1/r^3\rangle_p$ is an expectation value for the $p$ orbitals, $p_z(E)$ is the partial density of states ($p$-DOS) in the muffin-tin sphere around an atom, and $E_F$ is the Fermi energy. The integral $p_i^I(E_1)$ counts the number of $p_i$ electrons in a muffin-tin sphere with an energy less than $E_1$. In our region of the periodic table (Rh to Xe) the orbitals being filled are $4d$, $5s$, and $5p$. From these, the spherically symmetric $5s$ can be excluded to have a relation with $V_{zz}$. In the region where $4d$ is being filled (Rh to Cd) $V_{zz}$ hardly changes, and it changes much more from In to Xe where the $4d$ occupation is constant. This makes a significant $4d$ contribution to $V_{zz}$ unlikely. If, however $\Delta p(E_F)$ for X is plotted against $V_{zz}$, we see an excellent linear correlation with $V_{zz}$. This shows that the $p$ anisotropy $\Delta p(E_F)$ determines $V_{zz}$: if $p$ electrons accumulate in the $xy$ plane (where the four Eu neighbors are), then $V_{zz}$ is positive [Eq. (4)]. If the $p$ charge piles up preferentially along the $z$ direction, $V_{zz}$ is negative. This implies that we can use the $p$ anisotropy to interpret $V_{zz}$ in terms of the chemical bond, as done for instance in Ref. 21. We will heavily rely on this in the remainder of the paper.

We can go one step further in the interpretation if we plot $\Delta p(E)$, i.e., as a function of the energy $E$ [Fig. 4(c)]. For all X this gives a fairly similar function: a region of slightly negative values ($p_z$ excess) at low energies, followed by a region of strongly positive values ($p_{xy}$ excess), a steep decrease to strongly negative values ($p_z$ excess) and a rise to roughly zero (spherical symmetry). The main feature that distinguishes the different elements X is the position of the Fermi energy $E_F$, i.e., the filling of the $p$ band. The value $\Delta p(E=E_F)$ determines the actual $p$ anisotropy for that X, and hence determines $V_{zz}$. From Rh to Xe, the position of $E_F$ gradually moves from left to right through the $p$ anisotropy. For Rh and Pd, $E_F$ lies in the slightly negative region (small and negative $V_{zz}$). For Ag to Te it lies in the positive region, at ever larger values (growing and positive $V_{zz}$), and for I and Xe it went down the steep hill and lies in the negative region (large and negative $V_{zz}$). The heavier the X, the stronger the $p$ electrons are bound. This leads to a decreasing width of the $p$ band and hence to larger values of the anisotropy function. This explains why $V_{zz}$ can be (but not must be) larger for the heaviest X.

One can note a similarity between the trend of the squares in Fig. 3 and $V_{zz}$ on Ag $\rightarrow$Xe impurities in hcp Cd.[22–24] In the latter case the environment of the impurity is constant and built from Cd atoms only. In Fig. 3 the symmetry and distances of the environment are fixed, but the X neighbors change together with the probe. They still are $5sp$ elements, however, just as Cd is. Therefore the explanation given for the trend of impurities in Cd (Ref. 25 and Fig. 2 in Ref. 24) applies also here, and can be understood as an equivalent formulation of the idea described above and in Fig. 4.

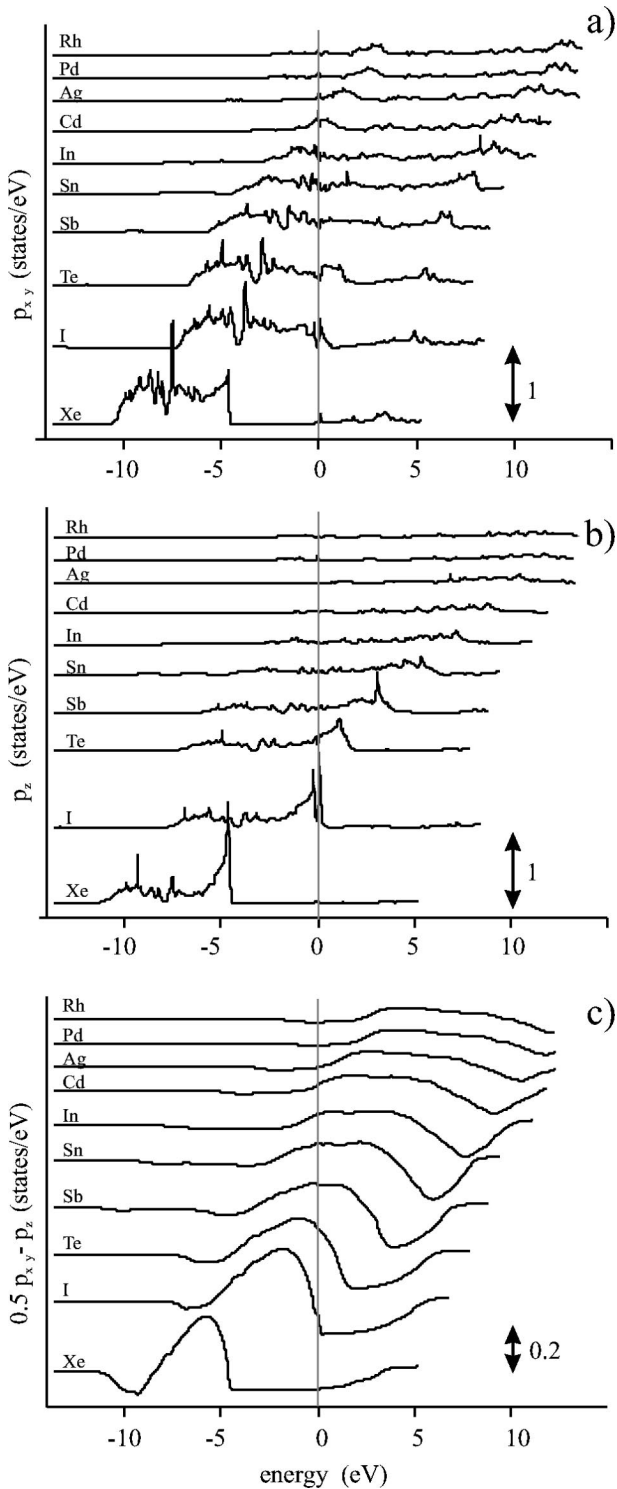

FIG. 4. (a) Partial $5p_{xy}$ DOS of X in Eu$X_3$. (b) Partial $5p_z$ DOS of X in Eu$X_3$. (c) $5p$ anisotropy $\Delta p(E)$ [as in Eq. (4) but as a function of energy] of X in Eu$X_3$. (For all pictures, the curves are vertically displaced for clarity. The leftmost part of each curve starts at zero on the vertical axis. The vertical axis is calibrated by the double arrow in each picture, which has the indicated length. The Fermi energy is at 0 eV.)

### B. $V_{zz}$ at Cd in Eu$X_3$

Working towards our actual problem, we now add Cd as an impurity to the series of (mostly hypothetical) Eu$X_3$ compounds. These are supercells with 16 atoms, Cd is at an $X$ site. In Fig. 3, $V_{zz}$ on Cd is shown (again using "regular" GGA). Here too there is one point (EuSn$_3$) that can be compared with experiment, and taking into account the arbitrary lattice constant and the absence of relaxation around the impurity, there is good agreement. The value of $V_{zz}$ on Cd is comparable to $V_{zz}$ on $X$ for Rh to Cd and then remains essentially constant at low positive values, except for I and Xe where it is either zero or large and positive. In Fig. 5(c), $\Delta p(E)$ is shown for Cd. For all cases, the shape of this function is quite similar to the one of Cd in EuCd$_3$. For the lighter $X$, $\Delta p(E)$ is rather broad and displays a maximum that is near 4 eV for $X=$Rh and moves closer to $E_F$ for heavier $X$. From $X=$Cd onwards, this maximum initially moves in the opposite direction, and then returns while $\Delta p(E)$ becomes narrower with steeper features. Apart from the beginning (Rh and Pd) and the end (I and Xe), the filling of the Cd $p$ band is fairly constant.

This behavior can be understood in detail by inspection of the DOS and consideration about the lattice structure and hybridization. We will not present this discussion here, as it would bring us too far from the main topic of the paper. Be it sufficient to mention that changes in $\Delta p(E_F)$ and the related changes in $V_{zz}$ are due to changes in Cd $p_{xy}$, while Cd $p_z$ remains fairly constant. At first sight this is unexpected, as it is Cd $p_z$ that interacts with the variable $X$, while Cd $p_{xy}$ interacts with the constant Eu (Fig. 1). However, for those $X$ where the deviation between $V_{zz}$ on Cd and on $X$ is largest, there are no $X$ $p_z$ states available at those energies where the dominant Cd $p_z$ weight is. This prevents direct interaction, and allows the indirect influence of $X$ $p_{xy}$ on Cd $p_{xy}$—mediated by Eu—to dominate.

### C. The role of $f$ electrons

Although DFT in its local density approximation (LDA) or GGA formulation provides an accurate description for many materials, it fails in a few situations. One example is when correlation effects become important, as is the case in lanthanides. Strong on-site Coulomb repulsion splits the $4f$ DOS.[26,27] For a rare earth element $R=[\text{Xe}]4f^n 5s^2$ with $n$ nominal $f$ electrons, an integer number of at least $n-1$ $f$ electrons are localized at the atomic site and do not participate in the bonding. In the $f$ DOS they yield a sharp peak well below (15 eV) the Fermi energy [Fig. 6(a)]. If the remaining $f$ electron is localized too, only two band electrons ($5s^2$) remain that are chemically active. In this case, the rare earth is said to be divalent. In such cases often an unoccupied $f$-electron peak is found in the DOS a few eV above $E_F$. In other cases, however, the remaining $f$-electron is bandlike too. It then participates in the bonding, and is seen in the DOS as a sharp $f$-electron peak that straddles the Fermi energy [Fig. 6(b)]. This single $f$ electron hybridizes with the rare earth $5s$ and $5d$ states, which therefore acquire some $f$ character. The filled part of the $f$-electron peak is left with less than one electron. Such a situation is labeled as trivalent, as three electrons are chemically active now. If LDA or GGA are used, the strong correlations between the $f$ electrons are largely missed: all $nf$ electrons are treated as band electrons, and consequently they are all found in a single, unsplit peak at $E_F$.[28–30] DFT calculations that go beyond the LDA/GGA level can to some degree improve on the treatment of correlation: LDA+$U$,[28] self-interaction correction (SIC),[27,30–34] open core calculations,[28,29] etc. Strange *et al.*[31] showed that with LDA+SIC a divalent or trivalent situation can be imposed on a rare earth atom in a compound (resulting in the correct DOS as given in Fig. 6), and that the lowest total energy is found for that valency that appears in nature (see also Refs. 27, 30, and 32–34). This proved that the DOS picture in Fig. 6 derived from experiment is correct, and we will use this picture from now on as the criterion to distinguish between valencies.

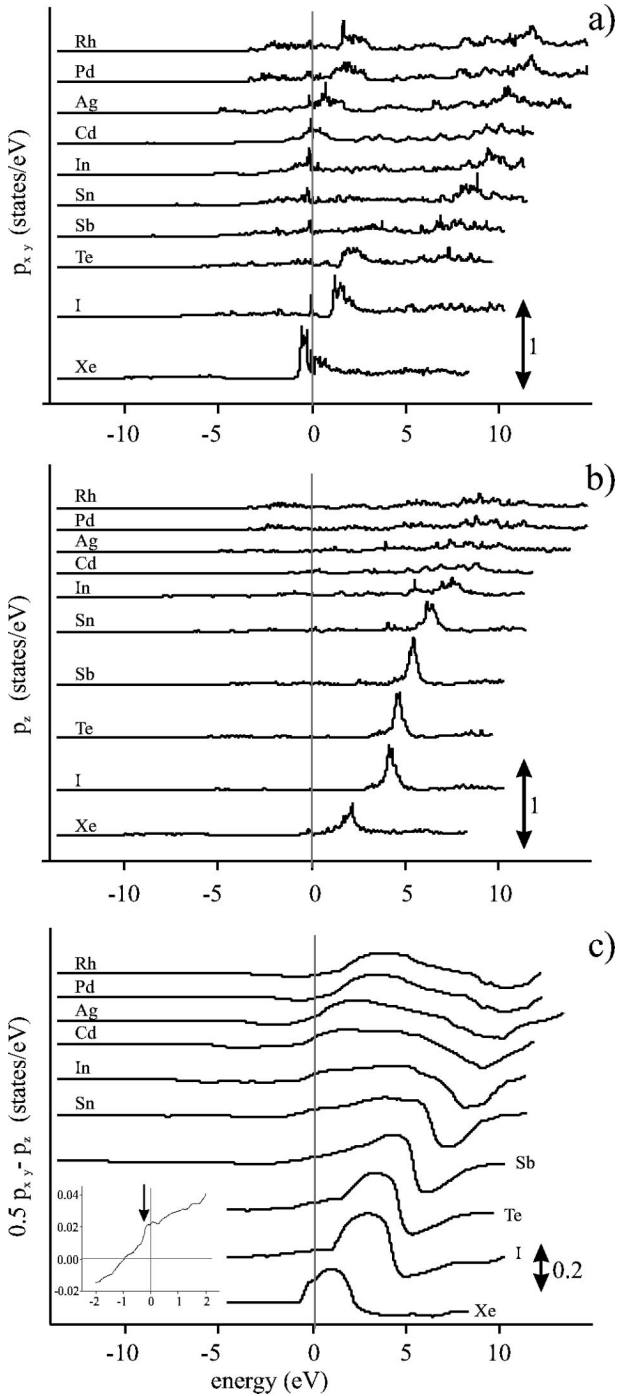

FIG. 5. (a) Partial $5p_{xy}$ DOS of Cd in Eu$X_3$. (b) Partial $5p_z$ DOS of Cd in Eu$X_3$. (c) $5p$ anisotropy $\Delta p(E)$ [as in Eq. (4) but as a function of energy] of Cd in Eu$X_3$. The inset shows a detail for EuSn$_3$. (For all graphs, the curves are vertically displaced for clarity. The leftmost part of each curve starts at zero on the vertical axis. The vertical axis is calibrated by the double arrow in each picture, which has the indicated length. The Fermi energy is at 0 eV.)

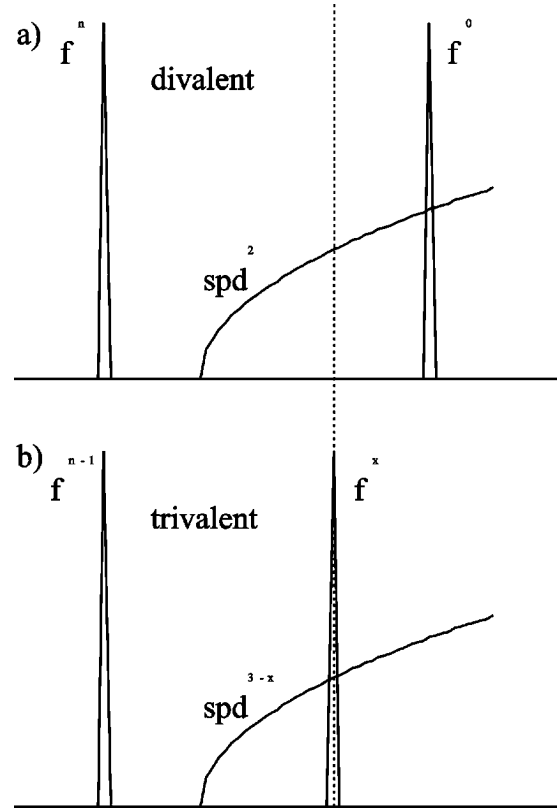

FIG. 6. Schematic picture of the partial DOS of a (a) divalent and (b) trivalent lanthanide in a compound, with $n$ nominal $f$-electrons per lanthanide atom. The vertical line indicates the Fermi energy.

We did not use LDA+SIC or LDA+$U$, but the less sophisticated "open core" scheme. The reason for this suboptimal choice—which was nevertheless sufficient for our purposes—was our need to calculate simultaneously $V_{zz}$, a quantity that is not implemented in many DFT codes. At the time this work was carried out, no code with LDA +$U$/LDA+SIC and EFG calculation was available [meanwhile, the new version of the WIEN code—WIEN2k (Ref.

TABLE I. $V_{zz}$ on Cd (16 atom supercel) and $X=$(In,Sn) (no supercell) in several $RX_3$ compounds, simulating trivalent and divalent situations with regular GGA and open core calculations, respectively.

| | $V_{zz}^{Cd}$ trivalent | $V_{zz}^{Cd}$ divalent | $V_{zz}^{In/Sn}$ trivalent | $V_{zz}^{In/Sn}$ divalent |
|---|---|---|---|---|
| SmIn$_3$ | 4.9 | 1.1 | 11.1 | 7.5 |
| EuIn$_3$ | 4.7 | 1.4 | 10.6 | 8.4 |
| GdIn$_3$ | 4.3 | 1.5 | 10.0 | 8.4 |
| SmSn$_3$ | 4.6 | 0.8 | 16.6 | 13.7 |
| EuSn$_3$ | 4.1 | 1.3 | 16.2 | 13.6 |
| GdSn$_3$ | 4.2 | 1.5 | 16.1 | 13.5 |
| EuCd$_3$ | 3.7 | 1.8 | | |

(35)—has LDA+$U$ included]. In an open core calculation, the $f$-electrons are removed from the valence bands, and are treated as atomic electrons. They cannot hybridize with the other valence $spd$ electrons any more and are perfectly localized. Such a situation is similar to the divalent case of Fig. 6(a), where the $f$ states have no effect on the occupied part of the broad bands below the Fermi energy. A "regular" (= no open core) LDA/GGA calculation, however, puts all $f$ electrons at the Fermi energy, which is similar to the trivalent situation of Fig. 6(b). In Table I, $V_{zz}$ on In or Sn and on Cd is given in SmSn$_3$, SmIn$_3$, EuSn$_3$, EuIn$_3$, GdSn$_3$, and GdIn$_3$ for the trivalent and divalent situations obtained with regular and open core calculations. This is compared with $V_{zz}$ on Cd in hypothetical EuCd$_3$. Both for Cd and In or Sn, $V_{zz}$ is consistently about $(2-3)\times 10^{21}$ V/m$^2$ lower in the divalent case. Using EuCd$_3$ as an example, we now examine the mechanism of this reduction. Figure 7(a) shows the total DOS near the Fermi energy for the trivalent and divalent cases. The only difference is the presence of the huge $f$ peak at the Fermi energy. In order to see how this $f$ peak influences $V_{zz}$ on Cd, the $p_{xy}$ DOS of Cd is compared for both valences in Fig. 7(b): in the trivalent case, a number of states has been moved from the unoccupied region just above the Fermi energy to the occupied region just below it. This indicates a strong $f$-$p$ hybridization in the $xy$ plane. Figure 7(c) shows that for the $z$ direction the $f$-$p$ hybridization is much less pronounced. The reason for this difference can be understood from Fig. 1: the $f$-carrying Eu atoms are in the local $xy$ plane that contains the Cd. The Cd $p_{xy}$ orbitals are pointing towards Eu, which is not true for Cd $p_z$. An increase of the number of occupied $5p_{xy}$ states while the number of occupied $5p_z$ states remains constant means that the $5p$ anisotropy function reaches a higher value at the Fermi energy [Fig. 7(d)], which corresponds to an increased $V_{zz}$ in the trivalent case.

This mechanism is summarized in Fig. 8: the effect of the presence of an $f$ peak on the $p_{xy}$ band is that it digs a hole at energies higher than the energy of the $f$ peak, and makes an extra $p_{xy}$ peak at the position of the $f$ peak. Depending on the position of $E_F$, the $p$ anisotropy and therefore $V_{zz}$ will be either unchanged ($E_F=E_1$), increased ($E_F=E_2$) or unchanged ($E_F=E_3$). The increase is of the order of magnitude of $3\times 10^{21}$ V/m$^2$. Looking at the criterion for di- or trivalency in Fig. 6, we see that $E_F=E_2$ corresponds to a trivalent case, while $E_F = E_1$ is the divalent one. This means we showed that and understood why the position of the $f$ peak lowers $V_{zz}$ with about $3 \times 10^{21}$ V/m$^2$ (order of magnitude) if the rare earth changes from a tri- to divalent state.

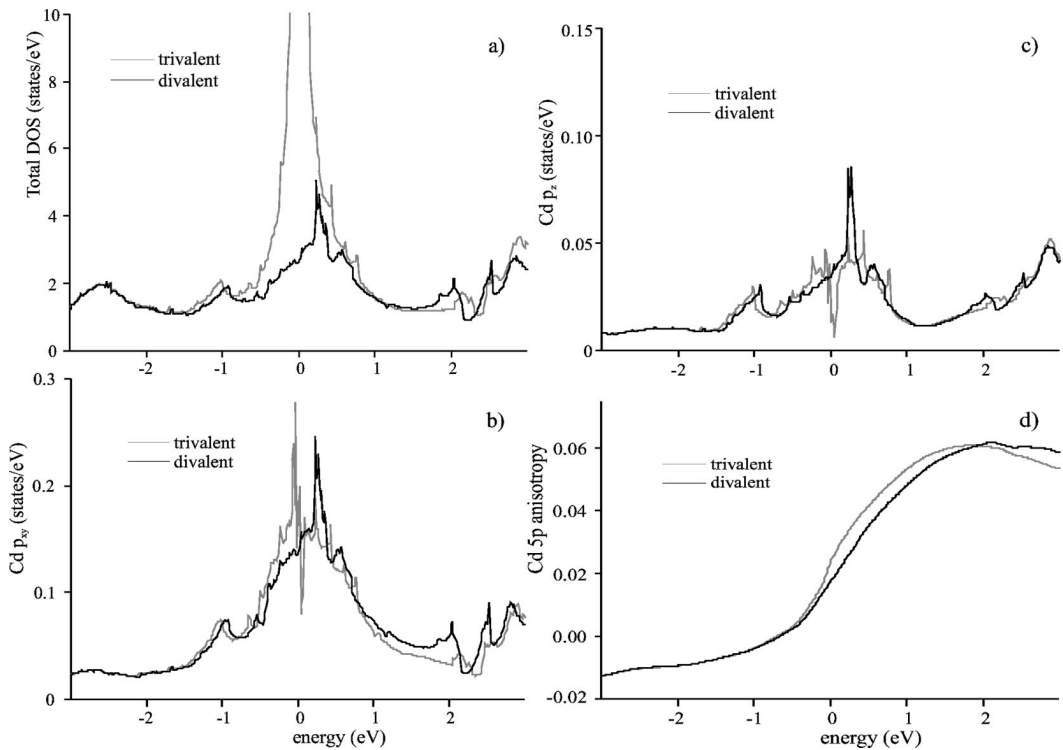

FIG. 7. All pictures are for EuCd$_3$. The solid line is for divalent Eu, the gray line for trivalent Eu. (a) Total DOS, (b) Cd $5p_{xy}$ DOS, (c) Cd $5p_z$ DOS, (d) Cd $5p$ anisotropy function.

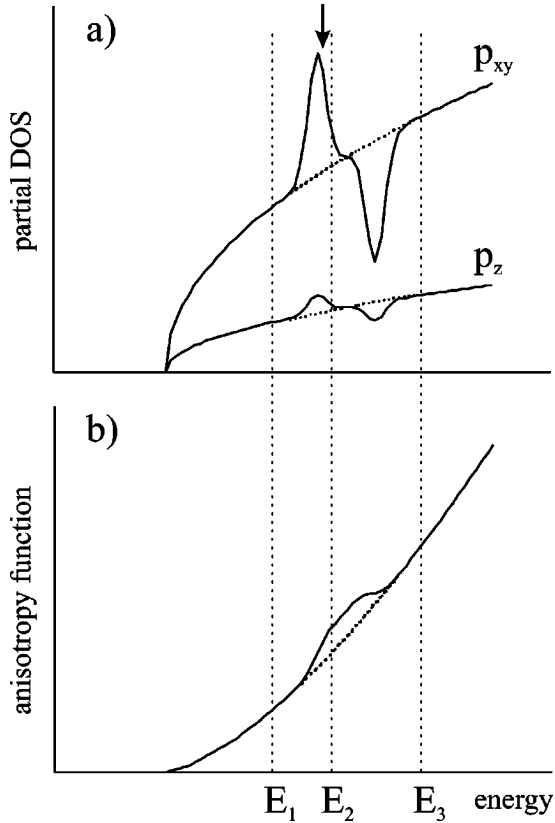

FIG. 8. Schematic presentation of the effect of the lanthanide $4f$ peak (arrow) on (a) the $5p_{xy}$ and $5p_z$ DOS of the Cd impurity and (b) the Cd $5p$ anisotropy function.

## V. DISCUSSION AND SUMMARY

In Sec. II D, four questions were raised. They have been explicitly or implicitly answered in Sec. IV. These answers will be made explicit and summarized now.

Question 1: Why is $V_{zz}$ at Cd in $R$In$_3$ and $R$Sn$_3$ strongly reduced for $R=$(Eu, Yb)? For trivalent rare earths, hybridization between rare earth $f$ states (in the $xy$ plane and near $E_F$) and Cd $p_{xy}$ states increases the Cd $p$ anisotropy [excess of $p_{xy}$ and hence $V_{zz}$ (Figs. 7 and 8)]. This increased $V_{zz}$ is the regular value of $(2-4) \times 10^{21}$ V/m$^2$ that is observed in Fig. 2(a). For divalent Eu and Yb the $f$ peak shifts a few eV higher to the unoccupied region. The increase of $V_{zz}$ is undone, and it drops by $3 \times 10^{21}$ V/m$^2$ (order of magnitude).

Question 2: Why is this not (or at least much less) the case for $V_{zz}$ at Sn? The same mechanism as for Cd exists for Sn, leading to the same *absolute* amount of reduction of $V_{zz}$ (Table I). This *absolute* effect is in contrast to the *relative* effect expected from the point charge model. Because $V_{zz}$ is much larger for Sn than for Cd, the effect is relatively smaller and easily obscured by other fluctuations in $V_{zz}$.

Question 3: Why is $V_{zz}$ at Cd in these compounds much smaller than $V_{zz}$ at Sn, and more generally, how can we understand the size of $V_{zz}$ at the 4/$mmm$ site in these compounds? The shape of the $p$ anisotropy as a function of the filling of the $5p$ band (Fig. 4 and Sec. IV A) is the answer. The Cd $p$ band is almost not filled, and the $p$ anisotropy is low in that region. The Sn $p$ band is filled with two electrons, and the anisotropy reaches a much higher value.

Question 4: Why is $V_{zz}$ at Cd in $R$Sn$_3$ half as large as in $R$In$_3$ if $R \neq$ (Eu,Yb)? The filling of the $p$ band involves changes of $V_{zz}$ in a range of $25 \times 10^{21}$ V/m$^2$. The difference between Cd in $R$Sn$_3$ and $R$In$_3$ is only $2 \times 10^{21}$ V/m$^2$, less than one-tenth of this. We saw that $V_{zz}$ of Cd tends to keep a similar value throughout the $RX_3$ series, and a difference of only $2 \times 10^{21}$ V/m$^2$ is "similar" on the scale of the $p$-band filling. We must therefore look for quantitative details that cannot be explained in our present approach, which is meant for discovering mechanisms. As seen in Fig. 3 and Table I, our calculated values for Cd in $R$Sn$_3$ are very similar to the values for EuIn$_3$. This can be due to the averaged lattice constant and to the details of the $p$ anisotropy for Cd. We cannot fully answer our fourth question, and can only conclude that more realistic modeling is needed here. A suggestion for a possible answer can be inferred from Fig. 5(c) (inset), where a significant drop in the $p$ anisotropy closely below the Fermi energy is seen. It would not be surprising if the Fermi energy in an accurate calculation is a little bit shifted to fall below this drop, which would considerably lower $V_{zz}$.

These answers were derived from an analysis of the shape and filling of the anisotropy function $\Delta p(E)$, which is calculated for the atom/nucleus of interest ($^{111}$Cd and $^{119}$Sn) and other related atoms in similar and maybe hypothetical compounds. The shape of the anisotropy function is correlated with the partial DOS of other atoms.

## VI. CONCLUSIONS AND OUTLOOK

We revisited electric-field gradient measurements that were interpreted by a PCM analysis to be a proof for the divalency of $R=$(Eu,Yb) in $RX_3$ ($X=$ In, Sn). The PCM seemed to work well for part of the problem, but not for other parts. Why? Because the apparently good results of the PCM were just good luck. The PCM predicts a relative drop of $V_{zz}$ if the valency changes. We showed it must be an absolute drop. By accident, the order of magnitude of this absolute drop is similar to the relative drop that the PCM predicts if for the effective charge $Z_R^{\text{eff}}$ the valency number is used (there is no compelling reason why this should be done). Our *ab initio* analysis shows how the drop of $V_{zz}$ can be related to a valency using the DOS. This gives insight in how the chemical bonds in these compounds are responsible for the value of $V_{zz}$.

In the past decades, work has been done on $V_{zz}$ in $RX_3$ compounds, where $R$ is an actinide instead of a rare earth.[36–39] The mechanism how to understand $V_{zz}$ as a function of $p$-band filling will remain valid for actinides, too. Moreover, for the light actinides up to Np all $5f$ electrons are itinerant (bandlike). No SIC is needed here, since LDA/GGA gives accurate results.[32,40] With the ideas presented here, one

VALENCY OF RARE EARTHS IN $R$In$_3$ AND $R$Sn$_3$: ... PHYSICAL REVIEW B **66**, 195103 (2002)cannot only obtain a deeper insight in the meaning of $V_{zz}$ for AuCu$_3$ actinide compounds, but even aim for quantitatively correct calculations. On a more general level, we expect that the method of EFG analysis we presented here can be useful in many more cases where physical information has to be deduced from EFG measurements.

## ACKNOWLEDGMENTS

Work in Leuven was financially supported by the Fund for Scientific Research–Flanders (FWO), Project No. G.0194.00 and the Inter-University Attraction Pole program IUAP P4/10. S.C. received support from the FWO. S.J.A., H.A., and R.S. acknowledge the financial support of the Isfahan University of Technology and also the Abdus Salam International Center for Theoretical Physics (ICTP).

## APPENDIX: THE POINT CHARGE MODEL

Not being able to calculate detailed electron densities (especially near the nucleus) in the 1970s, the wave functions were neglected. Instead, the charge distribution of the electrons that is continuous in reality was imagined to be concentrated as point charges on the nuclei, and each nucleus was attributed some effective charge $Z_n^{\text{eff}}$. The contribution to $V_{zz}$ of such a lattice of point charges can be calculated exactly by a sum over the entire lattice. For the case of the $4/mmm$ position in $RX_3$, this sum can be shown to converge to the expression[4,41–43]

$$V_{zz}^{\text{lat}} = \frac{e\, 8.67(Z_R^{\text{eff}} - Z_X^{\text{eff}})}{4\pi\epsilon_0 a_0^3}, \quad \text{(A1)}$$

with $a_0$ the lattice constant. Now the effect of the continuous wave functions—sensitive to the details of the chemical bond between an impurity $X_0$ and $(R,X)$—is reintroduced by two means. First, this EFG due to the lattice of point charges (which are external to the atom $X_0$) amplifies $V_{zz}^{\text{lat}}$ by a deformation of the electron cloud around $X_0$. This is expressed by the so-called Sternheimer antishielding factor $(1-\gamma_\infty^{X_0})$ that can be obtained by Hartree-Fock calculations and is available in tables for many atoms.[44] Values of $(1-\gamma_\infty^{X_0})$ are often large and positive, e.g., 30.27 for $X_0$=Cd and 23.34 for $X_0$=Sn. Second, conduction electrons that enter the region inside the atom $X_0$ create a similar enhancement with a factor $(1-k)$. This $k$ is called the electronic enhancement factor, and is in many cases found to be about 3.[45] Now $V_{zz}$ can be written as

$$V_{zz} = (1-k)(1-\gamma_\infty^{X_0})V_{zz}^{\text{lat}}. \quad \text{(A2)}$$

The electronic enhancement factor is a "leftover" parameter that absorbs all inaccuracies of the other parts of the model. It can be determined only by comparing a PCM calculation with a corresponding experiment. This makes the PCM absolutely unpredictive as far as the value of $V_{zz}$ is concerned.

Combining Eqs. (A1) and (A2), $V_{zz}$ can be expressed as a function of known quantities, with $k$ and $(Z_R^{\text{eff}} - Z_X^{\text{eff}})$ as parameters. The latter depends in a rather well-defined way on the host matrix only, while $k$ reflects in an uncontrollable way the chemistry between the host matrix and $X_0$.

The point charge model has a long tradition in the interpretation of electric-field gradients in solids. It is now well established that calculating the electric-field gradient from first principles is in all respects superior to the PCM (see, e.g., Refs. 8–13). Nevertheless, it appears hard to eradicate, and in some communities even still gets refined.[46]

---